%% file: main.tex
\documentclass[sigplan,screen]{acmart}

\AtBeginDocument{%
  \providecommand\BibTeX{{%
    \normalfont B\kern-0.5em{\scshape i\kern-0.25em b}\kern-0.8em\TeX}}}

\renewcommand\footnotetextcopyrightpermission[1]{}

\setcopyright{acmcopyright}\acmConference[ASPDAC '21]{26th Asia and South Pacific Design Automation Conference}{January 18--21, 2021}{Tokyo, Japan}

\input{packages}
\input{acronyms}

\input{shorthands}

\input{cmds}

\def\FIGDIR{./fig}

\begin{document}

\title{Dataflow-Architecture Co-Design for 2.5D DNN Accelerators using Wireless Network-on-Package}

\author{Robert Guirado}
\authornote{Both authors contributed equally to this research.}
\email{rguirado@ac.upc.edu}
\affiliation{%
  \institution{Universitat Polit\`{e}cnica de Catalunya}
  \city{Barcelona}
  \country{Spain}
}

\author{Hyoukjun Kwon}
\authornotemark[1]
\email{hyoukjun@gatech.edu}
\affiliation{%
  \institution{Georgia Institute of Technology}
  \city{Atlanta}
  \state{GA}
  \country{USA}
}

\author{Sergi Abadal}
\email{abadal@ac.upc.edu}
\affiliation{%
  \institution{Universitat Polit\`{e}cnica de Catalunya}
  \city{Barcelona}
  \country{Spain}
}
\author{Eduard Alarc\'{o}n}
\email{eduard.alarcon@upc.edu}
\affiliation{%
  \institution{Universitat Polit\`{e}cnica de Catalunya}
  \city{Barcelona}
  \country{Spain}
}

\author{Tushar Krishna}
\email{tushar@ece.gatech.edu}
\affiliation{%
  \institution{Georgia Institute of Technology}
  \city{Atlanta}
  \state{GA}
  \country{USA}
}



\renewcommand{\shortauthors}{R. Guirado and H. Kwon, et al.}

\begin{abstract}
\input{00-abstract} 
\end{abstract}


\maketitle

\input{01-introduction}
\input{02-background}

\input{03-motivation}

\input{04-wienna}

\input{05-evaluations}
\input{06-related_works}
\input{07-conclusion}
\input{08-acknowledgement}

\small{
\bibliographystyle{ACM-Reference-Format}
\bibliography{ref_nonabbr}
}
\end{document}

%% file: packages.tex
\newcommand{\ignore}[1]{}
\usepackage[normalem]{ulem}

\usepackage{comment}
\usepackage{caption}
\usepackage{multirow}
\usepackage{subfigure}
\usepackage{color, soul}
\usepackage[nolist,nohyperlinks]{acronym}
\usepackage{upgreek}
\usepackage{graphicx}
\usepackage{xspace}
\usepackage{epstopdf}

%% file: acronyms.tex
\begin{acronym}
\acro{WNoC}{Wireless Network-on-Chip}
\acro{OOK}{On-Off Keying}
\acro{BER}{Bit Error Rate}
\acro{MAC}{Medium Access Control}
\acro{FDMA}{Frequency Division Multiple Access}
\acro{TDMA}{Time Division Multiple Access}
\acro{CDMA}{Code Division Multiple Access}
\acro{CSMA}{Carrier Sensing Multiple Access}
\acro{mmWave}{millimeter-wave}
\acro{NoC}{Networks-on-Chip}
\acro{BPSK}{Binary Phase-Shift Keying}
\acro{QPSK}{Quadrature Phase-Shift Keying}
\acro{RF}{Radio Frequency}
\acro{CNN}{Convolutional Neural Networks}
\acro{PE}{Processing Elements}
\acro{WIENNA}{Wireless-Enabled communications in Neural Network Accelerators}
\end{acronym}

%% file: shorthands.tex
\newcommand{\ourwork}{\textsc{WIENNA}\xspace}

%% file: cmds.tex
\newcommand{\betterparagraph}[1]{\noindent{\textbf{#1. }}}

\newcommand{\insertWideFigure}[2]{

    \begin{figure*}[h]
    \setlength{\abovecaptionskip}{-1pt}
    \setlength{\belowcaptionskip}{-4pt}
        \centering
        \includegraphics[width=\textwidth]{\FIGDIR/#1.pdf}
	    \vspace{-3mm}
        \caption{\small #2}
    	\vspace{-2mm}
        \label{fig:#1}
    \end{figure*}
}

\newcommand{\squishlist}{
 \begin{list}{$\bullet$}
  { \setlength{\itemsep}{0pt}
     \setlength{\parsep}{3pt}
     \setlength{\topsep}{3pt}
     \setlength{\partopsep}{0pt}
     \setlength{\leftmargin}{1.5em}
     \setlength{\labelwidth}{1em}
     \setlength{\labelsep}{0.5em} } }

\newcommand{\squishlisttwo}{
 \begin{list}{$\bullet$}
  { \setlength{\itemsep}{0pt}
     \setlength{\parsep}{0pt}
    \setlength{\topsep}{0pt}
    \setlength{\partopsep}{0pt}
    \setlength{\leftmargin}{2em}
    \setlength{\labelwidth}{1.5em}
    \setlength{\labelsep}{0.5em} } }

\newcommand{\squishend}{
  \end{list}  }

\newcommand{\TK}[1]{\textcolor{red}{TK: #1}}

%% file: 00-abstract.tex
Deep neural network (DNN)
models continue to grow in size and complexity, demanding higher computational power to enable real-time inference. To efficiently deliver such computational demands, hardware accelerators are being developed and deployed across scales. 
This naturally requires an efficient scale-out mechanism for increasing compute density as required by the application. 2.5D integration over interposer has emerged as a promising solution, but as we show in this work, the limited interposer bandwidth and multiple hops in the Network-on-Package (NoP) can diminish the benefits of the approach. To cope with this challenge, we propose \ourwork, a wireless NoP-based 2.5D DNN accelerator. In \ourwork, the wireless NoP connects an array of DNN accelerator chiplets to the global buffer chiplet, providing high-bandwidth multicasting capabilities. Here, we also identify the dataflow style that most efficienty exploits the wireless NoP's high-bandwidth multicasting capability on each layer. With modest area and power overheads, \ourwork achieves 2.2X--5.1X higher throughput and 38.2\% lower energy than an interposer-based NoP design. 
\vspace{-4mm}

%% file: 01-introduction.tex
\section{Introduction}
\label{sec:intro}


Deep Neural Networks (DNN) are currently able to solve a myriad of tasks with superhuman accuracy~\cite{resnet,unet}. To achieve these outstanding results, DNNs have become larger and deeper, reaching upto billions of parameters. Due to the enormous amount of calculations, the hardware running DNN inference has to be extremely energy efficient, a goal that CPUs and even GPUs do not live up to easily. This has led to a fast development of specialized hardware. 

Research in DNN accelerators \cite{eyeriss2,maeri} is a bustling topic. DNNs exhibit plenty of data reuse and parallelism opportunities that can be exploited via custom memory hierarchies and a sea of processing elements (PEs), respectively. However, as DNN models continue to scale, the compute capabilities of DNN accelerators need to scale as well.

At a fundamental level, DNN accelerators can either be scaled up (i.e., adding more PEs) or scaled out (i.e., connecting multiple accelerator chips together). There is a natural limit to scale-up due to cost and yield issues. Scale-out is typically done via board-level integration which comes with overheads of high communication latency and low bandwidth.
However, the recent appeal of 2.5D integration of multiple {\em chiplets} interconnected via an interposer on the same package~\cite{chips_dac2019} (e.g., AMD's Ryzen CPU), offers opportunities for enabling efficient scale-out. This has proven effective in DNN accelerator domain via recent works~\cite{zimmer20190, shao2019simba}.

Unfortunately, as we identify in this work, DNN accelerator chiplets have an insatiable appetite for data to keep the PEs utilized, which is a challenge for interposers due to their limited bandwidth. Such a limitation originate from large microbumps ($\sim$40$\upmu$m~\cite{chips_dac2019}) at the I/O ports of each chiplet, which naturally reduces the bandwidth by orders of magnitude compared to the nm pitch wires within the chiplet. The limited I/O bandwidth also causes chiplets to be typically connected to its neighbours only, hence making the average number of intermediate chiplets (hops) between source and destination to increase with the chiplet count. This introduces significant delay to data communication.

To address these challenges, for the first time, the opportunities provided by integrated wireless technology \cite{Cheema2013, Yu2014, abadal2020graphenebased, replica} for the data distribution in 2.5D DNN accelerators are explored. We show that wireless links can (i) provide higher bandwidth than electrical interposers and (ii) naturally support broadcast. We then identify dataflow styles that use these features to exploit the reuse opportunities across accelerator chiplets.



This main contribution is WIENNA, \emph{WIreless-Enabled communications in Neural Network Accelerators}, a 2.5D on-package scale-out accelerator architecture that leverages single-cycle wireless broadcasts to distribute weights and inputs, depending on the partitioning (i.e., dataflow) strategy. We implement three partitioning strategies (batch, filter and activation) for the 
scale-out architecture, leveraging the parallelism in DNNs across these three dimensions. We evaluate an instance of WIENNA 
with 
256 NVDLA~\cite{nvdla}-like accelerator chiplets, and observe 2.5-4.4$\times$ throughput improvement and average 38.2\% energy reduction over a baseline 2.5D accelerator that uses an electrical-only interposer.

%% file: 02-background.tex
\vspace{-2mm}
\section{Background and Related Work}
\label{sec:background}

\label{sec:dnnaccelerators}
\betterparagraph{DNN Accelerators} A DNN accelerator is specialized hardware for running DNN algorithms, which provides higher throughput and energy-efficiency than GPUs and CPUs via parallelism and dedicated circuitry. The abstract architecture of most DNN accelerators~\cite{eyeriss2, flexflow, maeri} consists of an off-chip memory, a global shared memory, and an array of PEs connected via a Network-on-Chip (NoC). 

When mapping the targeted DNN into the PEs, one can apply different strategies or \textit{dataflows}. Dataflows consist of three key components: loop ordering, tiling, and parallelization, which define how the dimensions are distributed across the PEs to run in parallel. The dataflow space is huge and its exploration is an active research topic~\cite{maestro, Gao2017tetris, zhao2019mrna, understanding}, as it directly determines the amount of data reuse and movement.

DNN accelerators involve three phases of communication~\cite{maeri} handled by the NoC. (1) \textit{Distribution} is a few-to-many flow that involves sending inputs and weights to the PEs via unicasts or multicasts or broadcasts depending on the partitioning strategy of the dataflow. (2) \textit{Local} forwarding involves data reuse within the PE array via inter-PE input/weight/partial-sum forwarding. (3) \textit{Collection} involves writing the final outputs back to the global memory.

\label{sec:interposer}
\betterparagraph{2.5D Integration} 2.5D systems refer to the integration of multiple discrete {\em chiplets} within a package, either over a silicon interposer~\cite{chips_dac2019} or other technologies. 
2.5D integration is promising due to higher yield (smaller chiplets have better yields than large monolithic dies), reusability (chipletization can enable plug-and-play designs, enabling IP reusability), and cost (it is more cost-effective to integrate multiple chiplets than design a complex monolithic die).
It has been shown to be effective for DNN accelerators~\cite{zimmer20190, shao2019simba}.

A silicon interposer is effectively a large chip upon which other smaller chiplets can be stacked.
Thus, wires on the interposer can operate at the same latency as on-chip wires. However, the micro-bumps used to connect
the chiplet to the interposer are $\sim$40$\upmu$m in state-of-the-art TSMC processes~\cite{chips_dac2019}.
This is much finer than board-level C4 bumps ($\sim$180$\upmu$m), but much wider than that of NoCs. This limits the number of microbumps that fit over the chiplet area and, thus, limits the effective inter-chiplet bandwidth. 

In a scale-out DNN accelerator, limited interposer bandwidth slows down reads and writes from the global SRAM. While collection (i.e., write) can be hidden behind compute delay, distribution (i.e., read) is in the critical path~\cite{shao2019simba}. Provisioning wires to enable fast distribution can be prohibitive and, even then, multiple hops are unavoidable. Enhancing this bandwidth is the aim of this work.

\begin{figure}[!t]
\centering
\includegraphics[width=0.95\columnwidth]{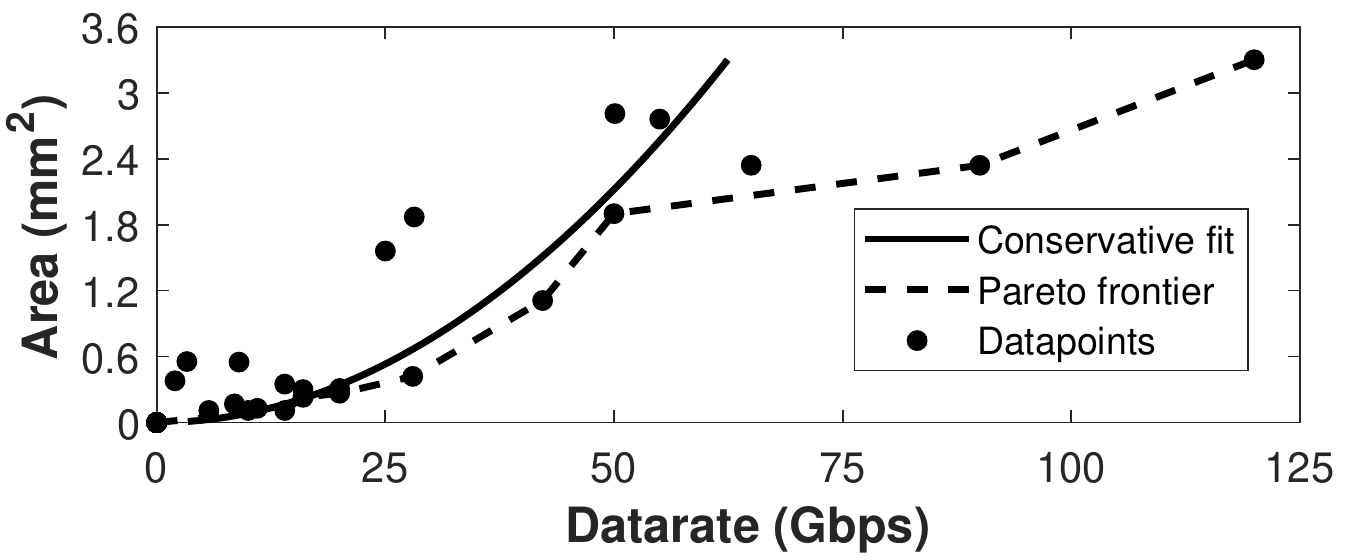}
\includegraphics[width=0.95\columnwidth]{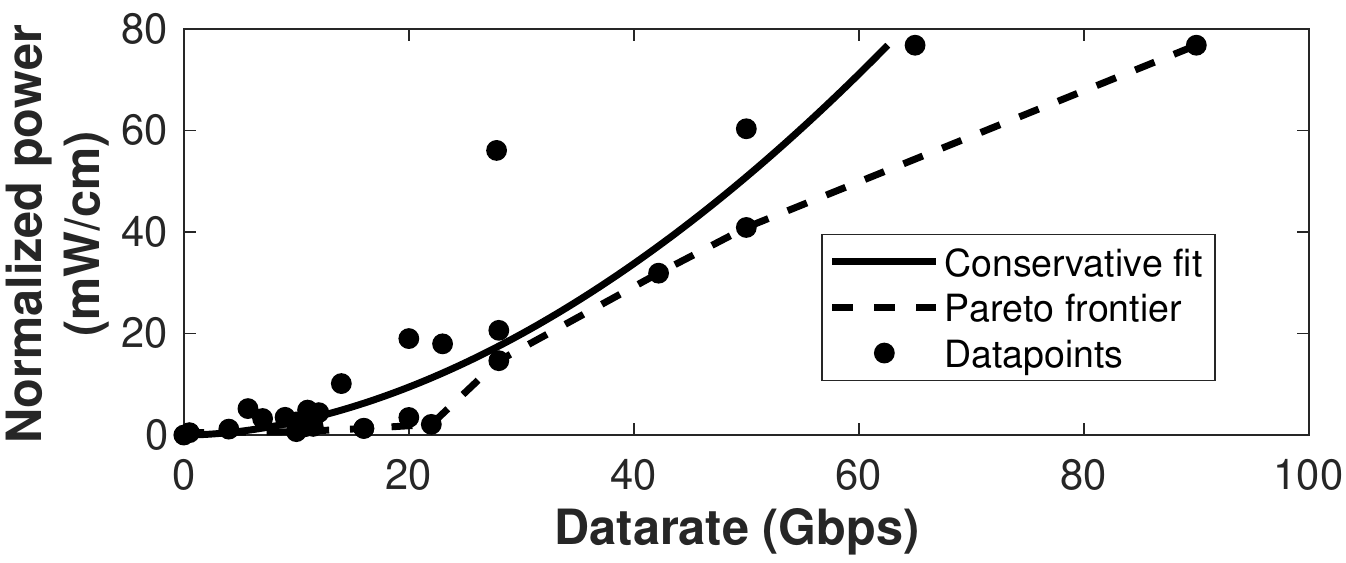} 
\vspace{-0.4cm}
\caption{\small Transceiver area and power as functions of the datarate for \cite{Tasolamprou2019, Yu2014, Tokgoz2018} and references therein. Power is normalized to transmission range and to 10\textsuperscript{-9} error rate. Energy per bit can be obtained dividing the power by the datarate. 
}
\label{fig:area_dr}\label{fig:power_dr}
\vspace{-0.8cm}
\end{figure}

\betterparagraph{Wireless Network-on-Package}
\label{sec:wnoc}
Fully integrated on-chip antennas \cite{Cheema2013} and transceivers (TRXs) \cite{Yu2014, Tokgoz2018} have appeared recently enabling short-range wireless communication upto 100 Gb/s and with low resource overheads. \autoref{fig:power_dr} shows how the area and power scale with the datarate, based on the analysis of 70+ short-range TRXs with different modulations and technologies \cite{Tasolamprou2019, Yu2014, Tokgoz2018}. 

Wireless NoPs are among the applications for this technology. In a wireless NoP, processor or memory chiplets are provided with antennas and TRXs that are used to communicate within the chiplet, or to other chiplets, using the system package as the propagation medium. This in-package channel is static and controlled and, thus, it can be optimized. In \cite{Timoneda2018ADAPT}, it is shown that system-wide attenuation below 30 dB is achievable. These figures are compatible with the 65-nm CMOS TRX specs from \cite{Yu2014}: 48 Gb/s, 1.95 pJ/bit at 25mm distance with error rates below 10\textsuperscript{-12}, and 0.8 mm\textsuperscript{2} of area. 

By not needing to lay down wires between TRXs, wireless NoP offers scalable broadcast support and low latency across the system. It is scalable because additional chiplets only need to incorporate a TRX, and not extra off-chip wiring, to participate in the communication. The bandwidth is high because it is not limited by I/O pin constraints, and latency is low as transfers bypass intermediate hops. Wireless NoP also allows to dynamically change the topology via reconfigurable medium access control or network protocols \cite{Shamim2017}, as receivers can decide at run-time whether to process incoming transfers. 

\insertWideFigure{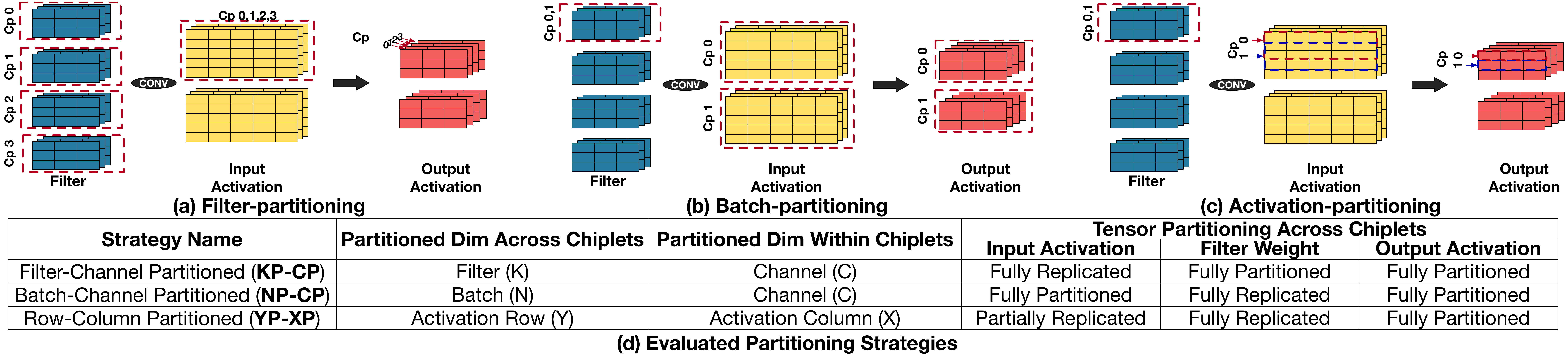}{Three tensor partitioning strategies across chiplets (a,b,c). Cp refers to chiplet. Based on the strategies, we construct three strategies as shown in (d). The replicated tensors are broadcast, while the partitiond tensors are unicast using the distribution network.
}

\insertWideFigure{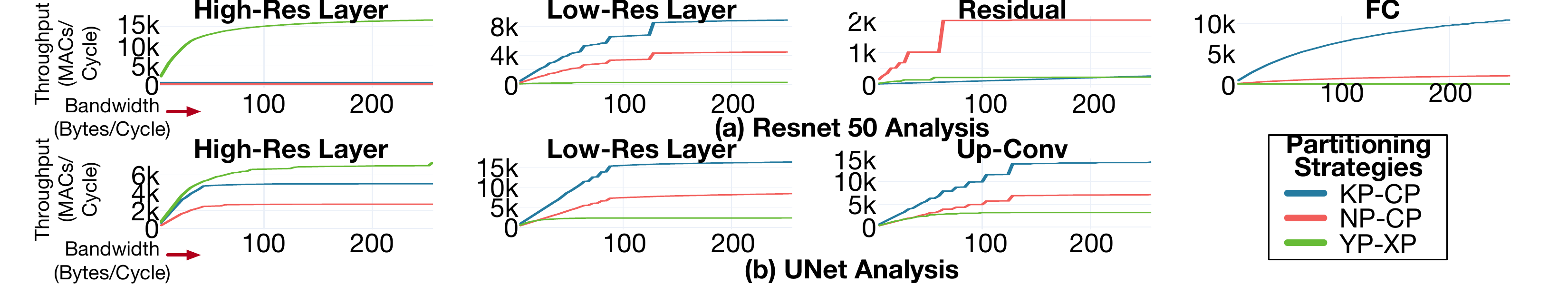}{The impact of bandwidth on throughput. We analyze a classification network, Resnet50~\cite{resnet}, and a segmentation network, UNet~\cite{unet} varying partitioning strategies. High-res and low-res layers indicate layers with larger/smaller activation height/width compared to the number of channels. Residual, FC, and Up-Conv indicate residual links, fully-connected layer, and up-scale convolutions.}

%% file: 03-motivation.tex
\vspace{-2mm}
\section{Motivation}
\label{sec:motivation}






To determine the design requirements for a 2.5D fabric for accelerator scale-out, we assess the impact of the data distribution bandwidth on the overall system throughput. \autoref{sec:methodology} details the simulation methodology and system parameters. 

We model a baseline 2.5D accelerator with 256 chiplets connected via a Mesh NoP. 
A global SRAM interfaces with DRAM on one end, 
and performs data distribution to the chiplets on the other. We implement three 
partitioning 
strategies~\cite{Gao2017tetris}, as shown in \autoref{fig:Partitioning_Strategies}.
Each chiplet is a 64-PE accelerator, implementing a dataflow optimized for that partitioning strategy.
We sweep the global SRAM read bandwith, 
and plot observed 
throughput in
~\autoref{fig:Motivation_BW_Impact}.
We run two state-of-the-art DNNs - ResNet~\cite{resnet} and UNet~\cite{unet} for image classification and image segmentation workloads.
Although we focus on two CNNs, they include a variety of layer operations and shapes, which can be a representative set of modern DNN layers, and significantly affect the performance and energy~\cite{maestro}.

We categorize layer types based on their operations and shapes in ~\autoref{table:layer_types} and, for each layer type, we plot the impact of bandwidth across the three partitioning strategies.

\input{Tables/LayersTable}



\betterparagraph{Observation I: Different layer types favor different partition strategies}
For instance, the high-resolution layers (i.e., input dim $>$ channel dim)
favor 
activation partitioning (i.e., YP-XP) across chiplets, where both inputs and weights can be broadcast.
Meanwhile, low-resolution layers and fully connected layers do not exhibit sufficient parallelism in activations, and 
favor filter partitioning (i.e., KP-CP) instead.

\betterparagraph{Observation II: different layer types saturate to peak throughput at different
bandwidth values}
High-res layers with 
YP-XP saturate to the peak best-case 
throughput of 16K MACs/cycle at 64 Bytes/cycle (i.e., 64 
unique inputs or weights delivered per cycle across the 256 chiplets) due to
effective bandwidth amplification due to
broadcasts of inputs and weights.
Low-res layers in ResNet saturate to 8K MACs/cycle with KP-CP beyond 128 Bytes/cycle.

\betterparagraph{Takeaways} Three takeaways from the above observations are that 
(i) the communication fabric for data distribution plays a key role in performance,
(ii) broadcast support and high-bandwidth are critical for scalability, and 
(iii) supporting adaptive partitioning strategies for each layer, rather than picking
a fixed one for all layers, is crucial for performance.

\betterparagraph{Challenges with Electrical NoPs}
As described in \autoref{sec:interposer}, bandwidth is the Achilles heel for interposer wires because it is limited by the microbump size.
For example, based on 55-$\upmu$m microbump size in one of the latest technologies~\cite{intel_aib}, only 21 wires can be placed over an edge of an accelerator~\cite{maeri} chiplet with 256 PEs.
According to that technology~\cite{intel_aib}, those 21 wires provide 42 Gbps bandwidth, 12.95$\times$ lower than the on-chip bandwidth in a chiplet from a recent work~\cite{shao2019simba}.
~\autoref{table:2.5DTech} compares these figures with other technologies.

Moreover, it is hard to design broadcast fabrics 
connecting hundreds of chiplets; broadcast will have to be supported via point-to-point forwarding, requiring multiple hops to deliver data to all chiplets, adding significant latency.
This also has a subtle side-effect in terms of synchronization since different chiplets will receive data 
at different times.

\betterparagraph{Promise of Wireless NoPs}
As discussed in \autoref{sec:background}, wireless NoPs are promising for 2.5D accelerators due to their broadcast support, independence from I/O pitch constraints, reconfigurability, and single-hop communication.

\begin{figure}[!t]
\setlength{\abovecaptionskip}{0pt}
\setlength{\belowcaptionskip}{-4pt}
\centering
\vspace{-0.1cm}
\includegraphics[width=0.95\columnwidth]{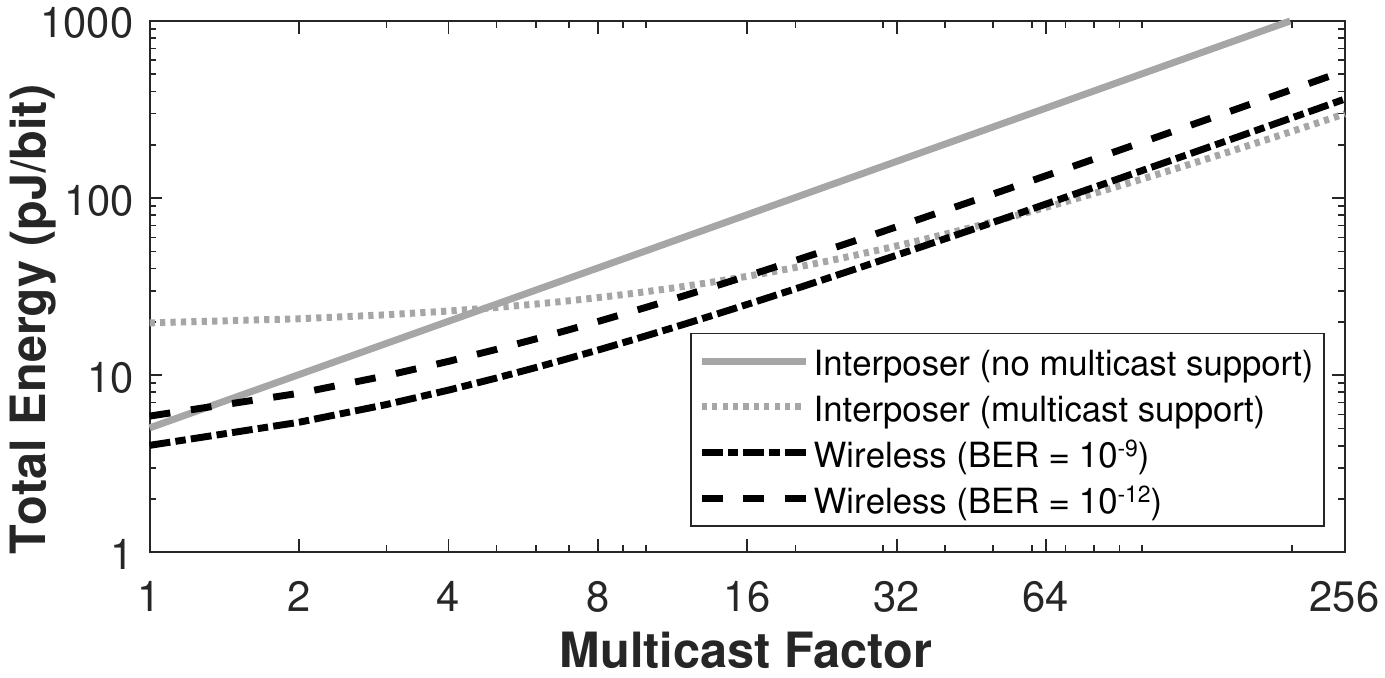}
\caption{\small Average per-bit energy of a multicast transmission in a silicon interposer with direct connections, mesh NoP with multicast support, and wireless NoP for two BER values.
}
\label{fig:multicastEnergy}
\vspace{-0.2cm}
\end{figure}

The single-hop communication is a key benefit in scale-out designs because the number of hops, which is a multiplier to latency and energy for interposer NoPs, increases with the communication fanout and the number of chiplets. Therefore, although some technologies provide higher bandwidth and lower energy per bit for single hop, as the number of chiplets or broadcast transmissions increase, the efficiency of wireless NoP surpasses other technologies. This is illustrated in~\autoref{table:2.5DTech} and~\autoref{fig:multicastEnergy}. This presents a co-design opportunity to design dataflows with multicast to leverage wireless, as we describe for our 2.5D accelerator next. 


\input{Tables/interconnectTechnologyTable.tex}

%% file: Tables/LayersTable.tex
\begin{table}[t]
\centering
\setlength{\abovecaptionskip}{0pt}
\setlength{\belowcaptionskip}{0pt}
\scriptsize
  
\begin{tabular} {|p{1.2cm}|p{6.5cm}|}
\hline
\textbf{Layer} & \textbf{Description} \\
\hline

High-res & CON2D layer with less channels than width of input activation\\
\hline

Low-res & CON2D layer with more channels than width of input activation \\
\hline

Residual & Skip connections \cite{resnet} \\
\hline

Fully-conn. & GEMM layer present in CNNs, MLPs, RNNs, and so on\\
\hline

UpCONV & Variant of CONV2D that increases the resolution of activation\\
\hline

 \end{tabular}
  \caption{\small Layer types}
  \label{table:layer_types}
\vspace{-5mm}  
\end{table}

%% file: Tables/interconnectTechnologyTable.tex
\begin{table}[t]
\centering
\setlength{\abovecaptionskip}{0pt}
\setlength{\belowcaptionskip}{0pt}
\scriptsize

\begin{tabular} {|p{2.5cm}|p{0.5cm}|p{0.5cm}|p{1.0cm}|p{0.5cm}|p{0.5cm}|}
\hline
\multicolumn{1}{|@{~} c @{~}|}{\textbf{Technology}}
& \multicolumn{1}{|@{~} c @{~}|}{\begin{tabular}{c} \textbf{Node} \\ \textbf{(nm)} \end{tabular}} 
& \multicolumn{1}{|@{~} c @{~}|}{\begin{tabular}{c} \textbf{BWD}\end{tabular}} 
& \multicolumn{1}{|@{~} c @{~}|}{\begin{tabular}{c} \textbf{Energy} \\ \textbf{(pJ/bit)} \end{tabular}}
& \multicolumn{1}{|@{} c @{}|}{\textbf{\begin{tabular}{c} \textbf{LL} \\ \textbf{(mm)} \end{tabular}}}
& \multicolumn{1}{|@{~} c @{~}|}{\begin{tabular}{c} \textbf{Avg \#} \\ \textbf{of Hops} \end{tabular}}
\\ 
\hline
\multicolumn{1}{|@{~} c @{~}|}{Silicon Interposer~\cite{dickson20128x}}
& 45
& 450
& 5.3
& 40
& O($\sqrt{N_{C}}$) 
\\ 
\hline
\multicolumn{1}{|@{~} c @{~}|}{Silicon Interposer~\cite{shao2019simba}}
& 16
& 80
& 0.82-1.75 
& 6.5*
& O($\sqrt{N_{C}}$)
\\ 
\hline
\multicolumn{1}{|@{~} c @{~}|}{EMIB (AIB)~\cite{intel_aib}}
& 14
& 36.4
& 0.85 
& 3
& O($\sqrt{N_{C}}$)
\\ 
\hline
\multicolumn{1}{|@{~} c @{~}|}{Optical Interposer~\cite{zia2016chip}}
& 40
& 8000
& 4.23
& N/A
& O($\sqrt{N_{C}}$)
\\ 
\hline
\multicolumn{1}{|@{~} c @{~}|}{Wireless (unicast)**}
& 65
& 26.5
& 4.01 
& 40
& 1
\\ 
\hline
\multicolumn{1}{|@{~} c @{~}|}{Wireless (broadcast)**}
& 65
& 64$\sqrt{N_{C}}$
& 1.4$N_{C}$ 
& 40
& 1
\\ 
\hline
 \end{tabular}
  \caption{\small 2.5D interconnect technologies. BWD refers to bandwidth density in Gbps/mm. $N_{C}$ represents the number of chiplets. LL refers to link length in mm. *Estimated based on package and chiplet dimensions. **Estimated based on Fig. \ref{fig:power_dr}. 
}
  \label{table:2.5DTech}
\vspace{-4mm}  
\end{table}



%% file: 04-wienna.tex
\vspace{-0.3cm}
\section{WIENNA Architecture}
\label{sec:wienna}

\begin{figure}[!t]
\setlength{\abovecaptionskip}{-4pt}
\setlength{\belowcaptionskip}{-4pt}
\centering
\includegraphics[width=\columnwidth]{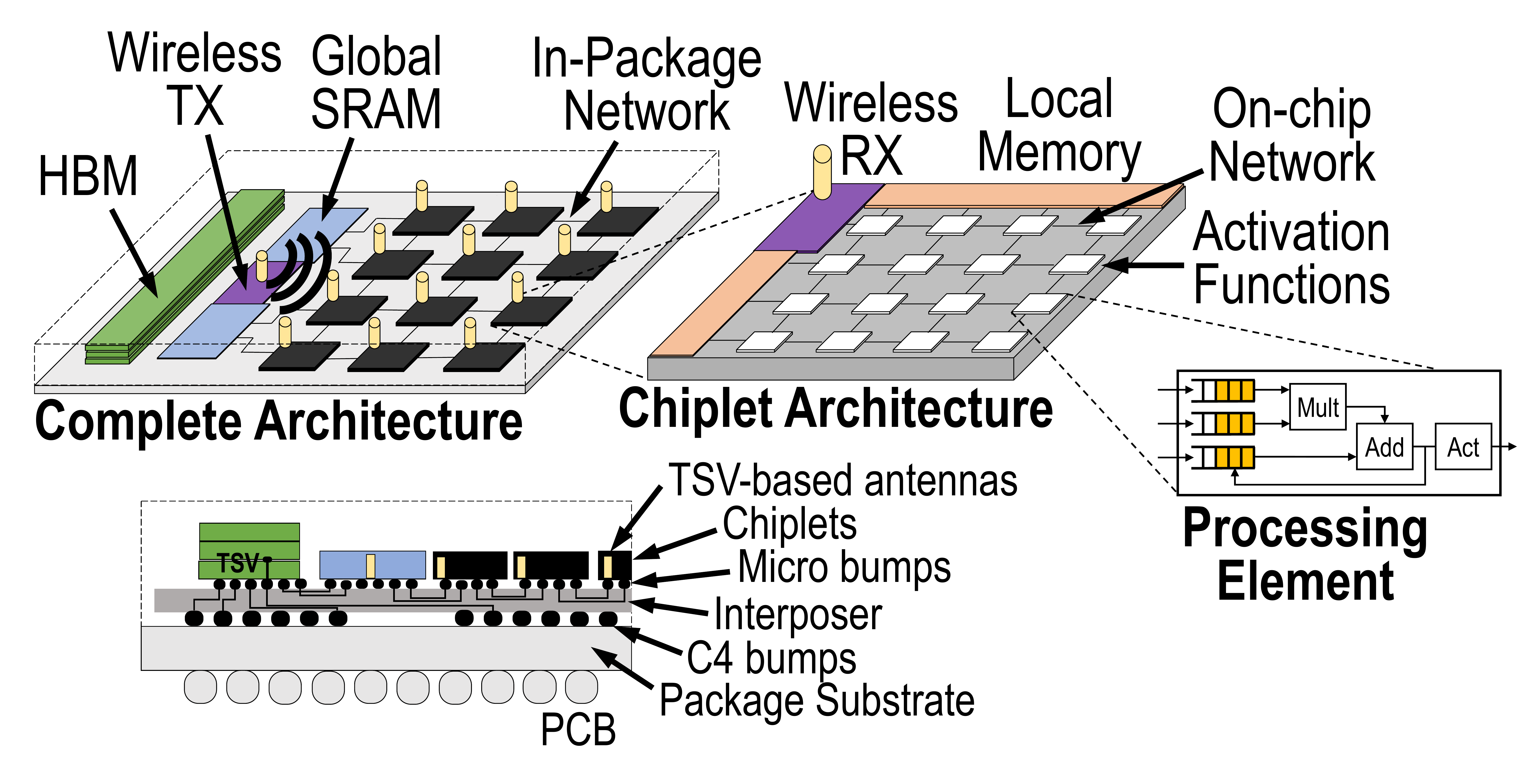}
\caption{\small Overview of the WIENNA architecture.}
\label{fig:system}
\vspace{-0.2cm}
\end{figure}

\autoref{fig:system} illustrates the WIENNA architecture. In essence, WIENNA contains a High Bandwidth Memory (HBM) that feeds a global SRAM memory chiplet, which is in turn connected to an array of accelerator chiplets. Each accelerator chiplet contains a local memory and an array of PEs, which are composed of a multiplier, an adder, as well as buffers that store inputs, weights, and outputs momentarily.

WIENNA implements a two-level hierarchy. On the one hand, the HBM, the global SRAM, and the array of chiplets follow a 2.5D integration scheme and are interconnected by means of a hybrid wired/wireless NoP. In the NoP, the wireless side is used for data distribution and the wired side for data collection. On the other hand, 
each chiplet implements its own internal microarchitecture. In this work, 
we use NVDLA~\cite{nvdla} and Shidiannao~\cite{shidiannao} style accelerators depending on the chosen workload partitioning strategy.
From a logical perspective, WIENNA's architecture allows us to partition the DNN dimensions across the chiplets via multiple mechanisms (see~\autoref{fig:Partitioning_Strategies}).
It also allows adaptive switching between these strategies for every layer of the DNN, building upon the reconfigurability of the wireless NoP.
%

\begin{figure}[!t]
\setlength{\abovecaptionskip}{0pt}
\setlength{\belowcaptionskip}{-4pt}
\centering
\includegraphics[width=\columnwidth]{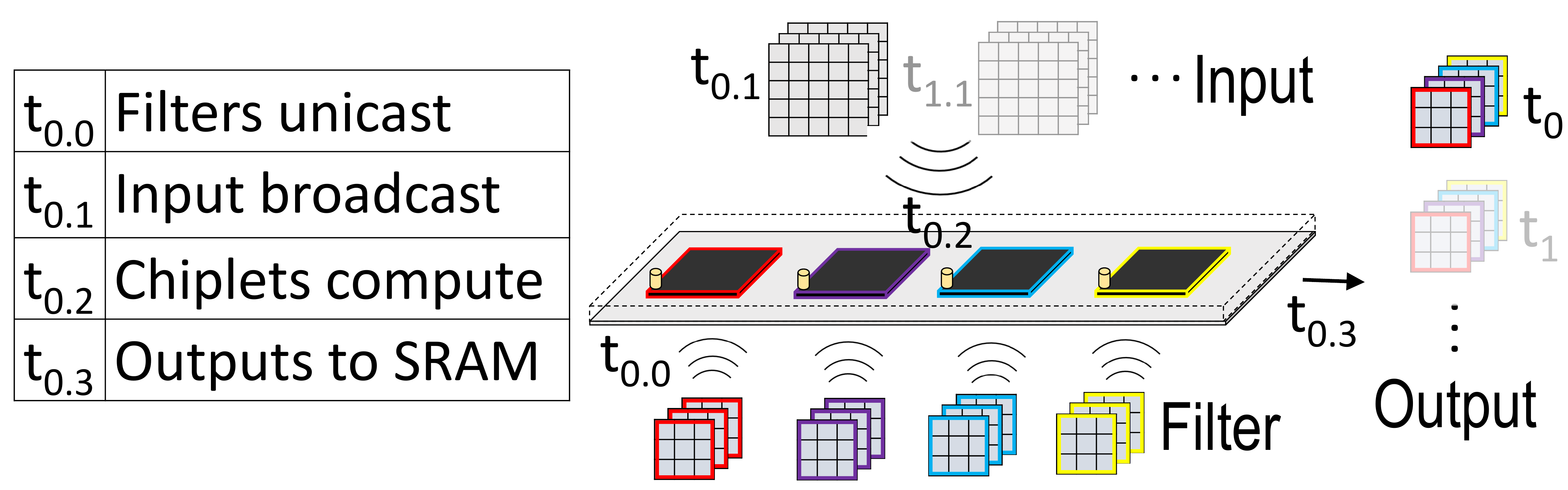}
\caption{\small WIENNA timeline for a filter partitioning example.}
\label{fig:timeline}
\vspace{-0.1cm}
\end{figure}

We describe a brief walkthrough example, showing how the KP-CP partition in~\autoref{fig:Partitioning_Strategies}(a) runs on WIENNA in \autoref{fig:timeline}.
In the example, the partitioned filters are first unicasted ($t_{0\_0}$) to each chiplet exploiting wide bandwidth of wireless networks.
Then, inputs are streamed by broadcasting ($t_{0\_1}$) one by one, exploiting the low-latency broadcasting of wireless networks.
Next ($t_{0\_2}$), each chiplet internally
distributes the inputs and weights following the intra-chiplet dataflow and computes the output activation.
Finally ($t_{0\_3}$), WIENNA utilizes the wired network to collect outputs from each chiplet. %

\betterparagraph{Interconnection Network} WIENNA uses a hybrid wired-wireless NoP. The wireless plane is \emph{only} used to distribute data from the SRAM to the chiplets, whereas the wired plane is used to collect the processed data 
back to the global SRAM. 

The wireless network is asymmetric (i.e., only distributes data) to keep it simple. If both distributions and reductions were to be performed through the wireless plane, full TRXs would be needed at each chiplet. Instead, WIENNA only requires a single TX located at the global SRAM and one RX per each chiplet. This avoids collisions completely, thus eliminating the need for a wireless arbiter and rendering flow and congestion control trivial because distributions are scheduled beforehand. The size and power of the TX and RX will depend on the required bandwidth (see \autoref{sec:wnoc}). 
Finally, note that TSV-based vertical monopoles \cite{Timoneda2018ADAPT} are assumed at both transmitting and receiving ends, as the use of such antennas reduces the losses at the chip plane.  


Besides wirelessly, the chiplets are also connected via a wired NoP through the interposer for output collection.
To combat pin limitations and wiring complexity, a mesh NoP is assumed \cite{shao2019simba,zimmer20190}.
We consider two design points with different bandwidth as listed in~\autoref{table:eval_setting}, to account for conservative and aggressive baselines of comparison.
In WIENNA, the wired NoP is only used for the collection phase.
In the baseline, it is used for both distribution and collection.

In summary, WIENNA's key feature is the proposal of an architecturally simple, but very powerful wireless NoP for low-cost, low-latency, and high-bandwidth data distribution from memory to the chiplets via unicast/broadcast.
WIENNA thus enables 2.5D chiplet scale-out.
Note that, while we evaluate a homogeneous chiplet array in this work, WIENNA makes no assumptions about the chiplet architecture and can thus accommodate heterogeneous combinations of chiplets with different architectures and networks.



\betterparagraph{Area and Power Overheads} To assess the implementation overhead of WIENNA,~\autoref{tab:res} shows the estimated area and power of an example WIENNA system with 256 chiplets and 64 PEs per chiplet at 65nm CMOS. We observe that the area overhead of a wireless RX is 16\% of a chiplet, which can be decreased when we employ a larger chiplet. Although wireless RX consumes 25\% of each chiplet's power, the delay benefits to be discussed in~\autoref{sec:results}, which is upto 5.1$\times$, compensate the power and eventually provide energy benefits (an average reduction of 38.2\%). Therefore, the overhead of WIENNA system is acceptable considering the benefits. 

\insertWideFigure{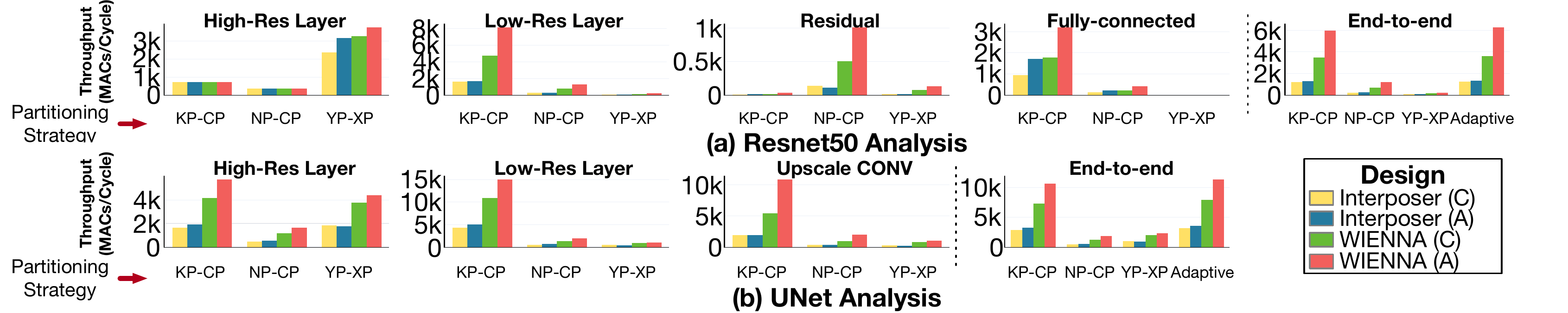}{Throughput analysis of conservative (C) and aggressive (A) designs of interposer- and WIENNA-based 2.5D accelerators.}

\insertWideFigure{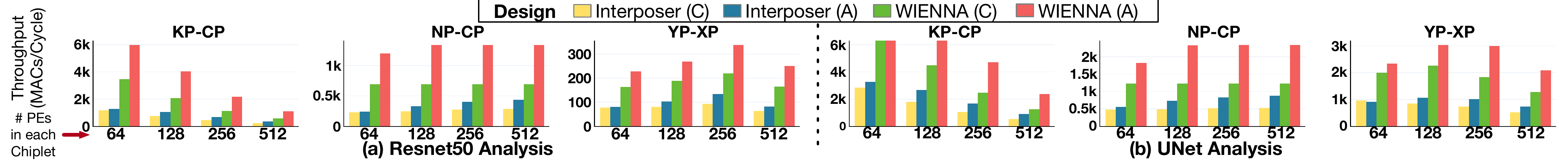}{Impact of cluster size for three partitioning strategies in (a) Resnet50 and (b) UNet.}

%% file: 05-evaluations.tex
\section{Evaluation}
\label{sec:evaluations}


\subsection{Methodology}
\label{sec:methodology}
We list up hardware parameters, workloads, partitioning strategy, and NoP characteristics in~\autoref{table:eval_setting}.
To compute the throughput, we use an open source DNN accelerator cost model, MAESTRO~\cite{maestro}, which is validated with average accuracy of 96.1\% against RTL simulation and measurements.
MAESTRO takes into consideration the latency, bandwidth, and multicasting characteristics of the different NoPs. To compute the energy of an electrical NoP, we compute the average number of hops multiplied by the per-hop energy from~\autoref{table:2.5DTech}. To estimate the energy of the wireless NoP, we select conservative (C) and aggressive (A) design points from~\autoref{fig:area_dr} at the required transmission rates. Note that~\autoref{fig:area_dr} assumes a single transmitter and receiver with a 50\%/50\% ratio, but this is actually a design choice. This allows to model the energy of both unicasts, where only the required receiver is active while others remain powered off to save energy; and broadcasts (multicasts), where all receivers (a set of receivers) are active.

\input{Tables/WIENNA-area-power.tex}

\subsection{Results}
\label{sec:results}
We compare the performance of interposer and WIENNA accelerators in~\autoref{fig:EvalResults} and~\autoref{fig:ClusterSz}. The energy is compared in~\autoref{fig:Eval_Energy} and the cause for their differences illustrated in~\autoref{fig:Eval_MultcastFactor}.

\input{Tables/EvalSettingTable.tex}

\insertWideFigure{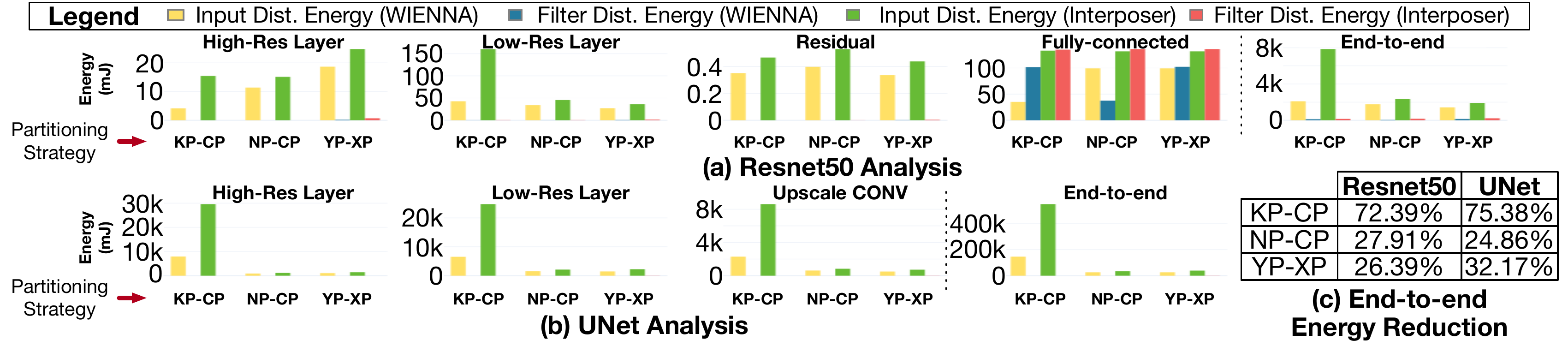}{Energy analysis of the distribution of input activations and filters from SRAM to chiplets 
in interposer- and WIENNA-based 2.5D accelerators. Inset  
(c) summarizes the end-to-end energy reduction by WIENNA compared to the interposer-based 2.5D accelerator.}

\insertWideFigure{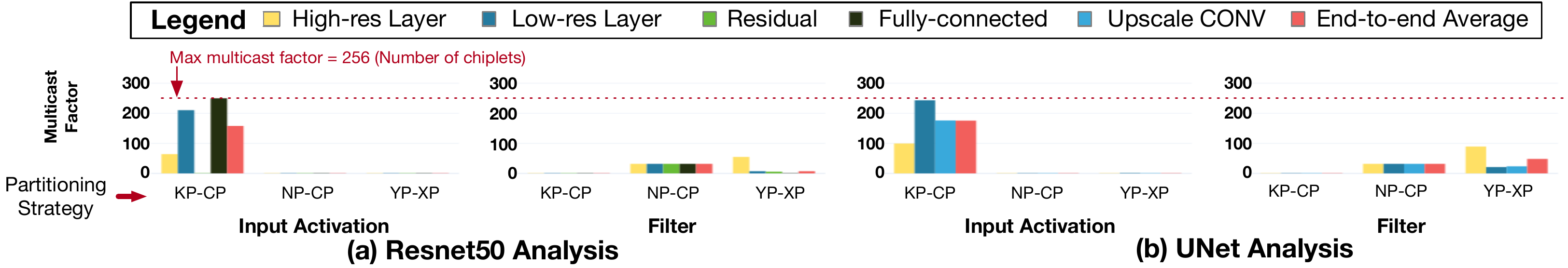}{Average multicast factor (number of received data across all the chiplets / number of sent data from global SRAM) for each layer types in the evaluated DNN models, (a) Resnet50 and (b) UNet. In this analysis, we apply the cluster size of 64, which results in 256 chiplets.}

\betterparagraph{Throughput Improvements}
\autoref{fig:EvalResults} presents the throughput analysis of WIENNA, from which we highlight several key results. First, WIENNA improves the end-to-end throughput by 2.7-5.1$\times$ on Resnet50 and 2.2-3.8$\times$ on UNet. Second, WIENNA can achieve better results than interposer with the same relative bandwidth. As listed in~\autoref{table:eval_setting}, aggressive interposer (interposer A) and conservative WIENNA (WIENNA C) have the same bandwidth, but WIENNA C provides 2.58$\times$ and 2.21$\times$ higher throughput than interposer A. The difference is based on the single-cycle broadcasting of the wireless NoP, much faster than the multi-hop wired baseline, which is critical in most partition methods as we observed in~\autoref{fig:Partitioning_Strategies} and later quantify in~\autoref{fig:Eval_MultcastFactor}. 
Third, the optimal partitioning strategy in terms of throughput and the impact of having higher bandwidth depends on the layer being processed. 
%
This is especially clear in ReNet50, where KP-CP works better in low-res and FC layers, NP-CP in residual layers, and YP-XP in high-res layers.
%
%
Based on this, in the end-to-end charts of~\autoref{fig:EvalResults}, we present results based on adaptive partitioning where we select the best strategy for each layer.
We find that adaptive partitioning improves throughput an extra 4.7\% and 9.1\% on Resnet50 and UNet, respectively, compared to keeping KP-CP across layers. 

We further evaluate throughput by studying the impact of varying the number of chiplets assuming a fixed total of 16384 PEs. 
For the interposer NoP, we adapt the number of hops to the resulting number of chiplets.
%
\autoref{fig:ClusterSz} shows results. 
Since the total number of PEs is fixed, less chiplets lead to more traffic per chiplet. 
%
%
Also, the utilization of chiplets and PEs in each chiplet depends on layer type and the partitioning strategy, as observed previously.
%
As a consequence, we do not observe a monotonic change of the throughput for all the cases and, thus, we conclude that the chiplet size is an important and optimizable design parameter. 
In any case, WIENNA is consistently faster and also more affected by the cluster sizes (77.5\% average difference from 64 to 512 PEs per chiplet) compared to interposer (62.5\%).
%
%

\betterparagraph{Energy Improvements}
The improvement in broadcast offered by WEINNA also affects the energy consumption. 
~\autoref{fig:Eval_Energy} compares the energy consumed in the distribution (from SRAM to chiplets) of input activations and filters in both systems. 
Across all the partitioning strategies and layers, WIENNA always reduces energy consumption (average of 38.2\%).
The energy saving is due to both the efficient multicast support via wireless NoP and the ample multicast opportunities of each partitioning strategy.
%
%
\autoref{fig:Eval_MultcastFactor} quantifies the multicast opportunities by plotting the multicast factor, which is the average number of destinations of each transfer from the global SRAM.
%
%
The multicast factor widely varies across partitioning strategies and layers, as they define the spatial reuse opportunities that determine the amount of multicast.
%
%
In general, we observe that the energy reduction provided by WIENNA is high when the multicast factor is high. A clear example is the KP-CP partitioning strategy leading to both the highest multicast factor in~\autoref{fig:Eval_MultcastFactor} and highest energy reduction in~\autoref{fig:Eval_Energy}.
%
%

%% file: Tables/WIENNA-area-power.tex
\begin{table}[t]
\centering
\scriptsize
\begin{tabular}{l| c c c| c c c}
\multicolumn{1}{l} {\textbf{Component}} & \multicolumn{3}{|c|}{\textbf{Area}} & \multicolumn{3}{c}{\textbf{Power}} \\
\quad \quad \textbf{Sub-element} & (mm\textsuperscript{2}) & (\%) & (\%) & (mW) & (\%) & (\%) \\  \hline 
Chiplets (256$\times$) &  1646 & 97  &   & 89600 & 89  & \\ 
\hline
\quad\quad PEs (64$\times$) + Mem  & 5 &   & 78 & 90 &   & 26 \\
\quad\quad Wireless RX & 1 &   & 16 & 90 &    &  25 \\
\quad\quad Collection NoP Router & 0.43 &   & 6 & 170 &    &  49 \\
\hline
Memory (1$\times$) &  53 & 3   &   & 10167 & 11   &    \\
\hline
\quad\quad Global SRAM & 51 &    & 96 & 10000 &    & 99\\
\quad\quad Wireless TX & 2   &  & 4 & 167 &    &  1\\
\hline
Total   & 1699 & 100   &    & 99767 & 100   &   \\ 
\hline
\end{tabular}
\caption{{\small WIENNA area and power breakdown for 256 chiplets, each with 64 PEs (16K MACs). The global SRAM is 13MB. The PE and SRAM data are based on Eyeriss~\cite{eyeriss2}. Wireless TX and RX are estimated from \autoref{sec:wnoc}, based on 10\textsuperscript{-9} BER. All data at 65-nm CMOS.} 
}
\vspace{-0.8cm}
\label{tab:res}
\end{table}

%% file: Tables/EvalSettingTable.tex
\begin{table}[t]
\centering
\setlength{\abovecaptionskip}{0pt}
\setlength{\belowcaptionskip}{0pt}
\scriptsize
  
\begin{tabular} {|p{2.3cm}|c|}
\hline
\multicolumn{1}{|@{~} r @{~}|}{\textbf{Total Number of PEs}}
& 16384 \\
\hline

\multicolumn{1}{|@{~} r @{~}|}{\textbf{Global SRAM Size}}
& 13 MiB \\
\hline

\multicolumn{1}{|@{~} r @{~}|}{\textbf{Clock Frequency}}
& 500 MHz \\
\hline

\multicolumn{1}{|@{~} r @{~}|}{\textbf{Number of Chiplets}}
& 32--1024 \\
\hline

\multicolumn{1}{|@{~} r @{~}|}{\textbf{PEs per Chiplet}}
& 64--512 \\
\hline

\multicolumn{1}{|@{~}r@{~}|}{\textbf{Workloads}}
& Resnet50 (Classification), UNet (Segmentation) \\
\hline

\multicolumn{1}{|@{~} r @{~}|}{\textbf{Partitioning Strategy}}
& KP-CP, NP-CP, and YP-CP \\
\hline

\multicolumn{1}{|@{~}r@{~}|}{\multirow {2}{*}{\textbf{Chiplet Architecture}}}
& KP-CP and NP-CP: NVDLA-like~\cite{nvdla} \\ 
& YP-XP: Shidiannao-like~\cite{shidiannao} \\
\hline

\multicolumn{1}{|@{~}r@{~}|}{\textbf{Interposer Bandwidth}}
& 8--16 Bytes/cycle/link (conservative--aggressive) \\ 
\hline

\multicolumn{1}{|@{~}r@{~}|}{\textbf{WIENNA Bandwidth}}
& 16--32 Bytes/cycle (conservative--aggressive) \\ 
\hline

\multicolumn{1}{|@{~}r@{~}|}{\textbf{Average Hops in NoP}}
& Interposer (mesh): $\sqrt{N_{chiplets}}/2$, WIENNA: 1 \\

\hline

\multicolumn{1}{|@{~}r@{~}|}{\textbf{Multicasting Capability}}
& Interposer: No, WIENNA: Yes \\
\hline

 \end{tabular}
  \caption{\small Evaluation settings.}
  \label{table:eval_setting}
\vspace{-3mm}  
\end{table}

%% file: 06-related_works.tex
\section{Related Works}
\vspace{-1mm}
\textbf{2.5D Chiplet Scaleout.}
DNN accelerator on-package scale-out is a nascent research area.
NVIDIA recently demonstrated a multi-chip-module accelerator~\cite{zimmer20190}.
WIENNA can augment such a design via the single-cycle wireless broadcast.
Gao {\emph et al.} proposed Tetris~\cite{Gao2017tetris}, a DNN accelerator
using a 3D memory exploiting wide bandwidth via TSV.
Tetris solves the bandwidth problem by leveraging TSVs to deliver data to every
compute tile. WIENNA solves the same problem for a 2.5D system via wireless NoP.
%
Simba~\cite{shao2019simba} taped out a 2.5D accelerator with 36K MAC units in 36
chiplets, which demonstrates the ability of 2.5D techonolgy to scale-out a DNN
accelerator.
However, the silicon interposer-based NoP among chiplets provide 10.88 $\times$
less bandwidth than on-chip bandwidth, which can be the system bottleneck 
depending on the workload and dataflow.
WIENNA can replace the NoP and provide higher bandwidth for higher performance.

\textbf{Wireless NoC.} The use of WNoCs for deep learning has been analyzed in \cite{Sinha2018,Choi2018}, but unlike WIENNA, the proposed architectures are single-chip and do not scale out. In \cite{Sinha2018}, a single dataflow is adapted to leverage the WNoC to broadcast weights in a CNN accelerator. The heterogeneous CPU/GPU platform from \cite{Choi2018} is not an accelerator, but is evaluated for deep learning workloads only. A multi-chip architecture enhanced with wireless interconnects is proposed in \cite{ascia2020}, but a single short-range TRX is taken as a baseline and only a single fixed dataflow is analyzed. WIENNA is the first scaleout-friendly architecture that leverages a wireless NiP, using concepts such as TSV antennas \cite{Pano2019}, asymmetric wireless multi-chip design \cite{ahmed2018asymmetric} and reconfigurable dataflows.

\vspace{-2mm}

%% file: 07-conclusion.tex
\vspace{-2mm}
\section{Conclusion}
\vspace{-1mm}
\label{sec:conclusion}
This paper proposes a new scalable design methodology of 2.5D DNN accelerators based on wireless technology.
We identify the required capabilities for the interconnect in 2.5D DNN accelerators and demonstrate that in those environments, wireless NoPs lead to 2.5-4.4$\times$ higher throughput and 38.2\% lower energy than interposer-based systems.

\vspace{-2mm}

%% file: 08-acknowledgement.tex
\section{Acknowledgements} 
\vspace{-1mm}
This work was supported by the European Commission under grant 863337 and NSF under Award OAC-1909900.
\vspace{-2mm}